\documentclass[a4paper,11pt]{article}
\usepackage{pos}
\usepackage{amsmath,bm}
\usepackage{graphicx}
\usepackage{slashed}
\usepackage{braket}
\usepackage{booktabs}
\usepackage{wasysym}

\def \nnb{\nonumber}

\newcommand{\order}[1]{\mathcal{O}\left(#1\right)}

\def\nnb{\nonumber}
\def\bea{\begin{eqnarray}}
\def\eea{\end{eqnarray}}

\def\be{\begin{equation}}
\def\ee{\end{equation}}

\title{
\vspace*{-7.6em}
\mbox{}\hfill \mbox{\small\sc SI-HEP-2022-07, P3H-22-040}\\
\vspace*{7.6em}
Some recent developments in nonleptonic $B$ decays}
\ShortTitle{Some recent developments in nonleptonic $B$ decays}

\author*[]{Tobias Huber}

\affiliation[]{Theoretische Physik 1, Naturwissenschaftlich-Technische Fakult\"at,
 Universit\"at Siegen, \\Walter-Flex-Stra{\ss}e 3, D-57068 Siegen, Germany}

\emailAdd{huber@physik.uni-siegen.de}

\abstract{I report on three recent topics from the field of two-body nonleptonic decays of $B_{(s)}$ mesons. The computation of two-loop NNLO corrections to the leading penguin amplitudes in QCD factorisation, puzzles that have emerged in colour-allowed tree-level decays to heavy-light final states, and a combination of QCD factorisation with $SU(3)$ flavour symmetry to estimate the size of weak-annihilation amplitudes.}

\FullConference{%
  11th International Workshop on the CKM Unitarity Triangle (CKM2021)\\
  22-26 November 2021\\
  The University of Melbourne, Australia
}

\begin{document}
\maketitle

\section{Introduction}

Nonleptonic decays of $B_{(s)}$ mesons play an essential role in the flavour-physics programmes at current and future colliders. The prospects for ever more precise measurements are therefore excellent, and that precision must be matched by theoretical predictions if we aim for improving our understanding of the mechanism of quark flavour mixing and CP violation, or even for pursuing the quest for indirect signals of new physics. On the theoretical side the bottleneck for precision is the computation of the hadronic matrix elements, where QCD effects from many different scales arise. Several approaches have been developed
to get a handle on the hadronic matrix elements, each having its virtues and drawbacks. {\emph{Factorisation frameworks}} such as PQCD~\cite{Keum:2000ph,Lu:2000em} or QCD factorization (QCDF)~\cite{Beneke:1999br,Beneke:2000ry,Beneke:2001ev} factorise short- from long-distance physics in the heavy quark limit. However, in their present use, they don't allow for the computation of sub-leading power corrections in the heavy-quark expansion from first field-theoretical principles (see~\cite{Khodjamirian:2005wn} for a computation of annihilation amplitudes with light-cone sum rules). {\emph{Flavour symmetries}} of the light quarks~\cite{Zeppenfeld:1980ex,Savage:1989ub,Gronau:1990ka} have the advantage of hardly requiring any assumption about the scales of the occurring QCD effects, and relate different decay channels to each other, thereby reducing the number of independent parameters. On the other hand, it is well-known that flavour $SU(3)$, $U$-spin and $V$-spin are severely broken by the strange-quark mass, and the lack of a rigorous implementation of flavour breaking can still be regarded as the main drawback of this approach. {\emph{Dalitz plot analyses}} are mostly applied to three-body decays, and very important for phenomenology. They are mostly data-driven, but also QCD-based predictions have been worked out in recent years~\cite{Krankl:2015fha,Klein:2017xti,Huber:2020pqb}. 

One obvious idea to get a better handle on nonleptonic decays is to combine the different approaches, with the goal of benefitting from their advantages, while at the same time minimizing the sensitivities to their individual drawbacks. One example is the so-called factorization-assisted topological amplitude approach~\cite{Li:2012cfa,Qin:2013tje,Zhou:2015jba,Wang:2017hxe,Jiang:2017zwr}, while combinations of factorization and flavour symmetries were studied e.g.\ in~\cite{Gronau:1995hn,Descotes-Genon:2006spp,Cheng:2011qh,Hsiao:2015iiu} and very recently in~\cite{Huber:2021cgk}. Also numerous studies of flavour symmetries in multibody final states exist (e.g.~\cite{Bhattacharya:2014eca,Bhattacharya:2015uua,Bediaga:2021okg}).

In this proceedings contribution, we report on three recent studies in the field of two-body nonleptonic $B_{(s)}$ decays. The two-loop ${\cal O}(\alpha_s^2)$ correction to the leading penguin amplitudes in QCD factorisation, puzzles that have emerged in colour-allowed tree-level decays to heavy-light final states, and a combination of QCD factorisation with $SU(3)$ flavour symmetry to estimate the size of weak annihilation amplitudes.

\section{Penguin amplitudes and direct CP asymmetries to ${\cal O}(\alpha_s^2)$}

\begin{figure}
\begin{center}
\includegraphics[scale=.33]{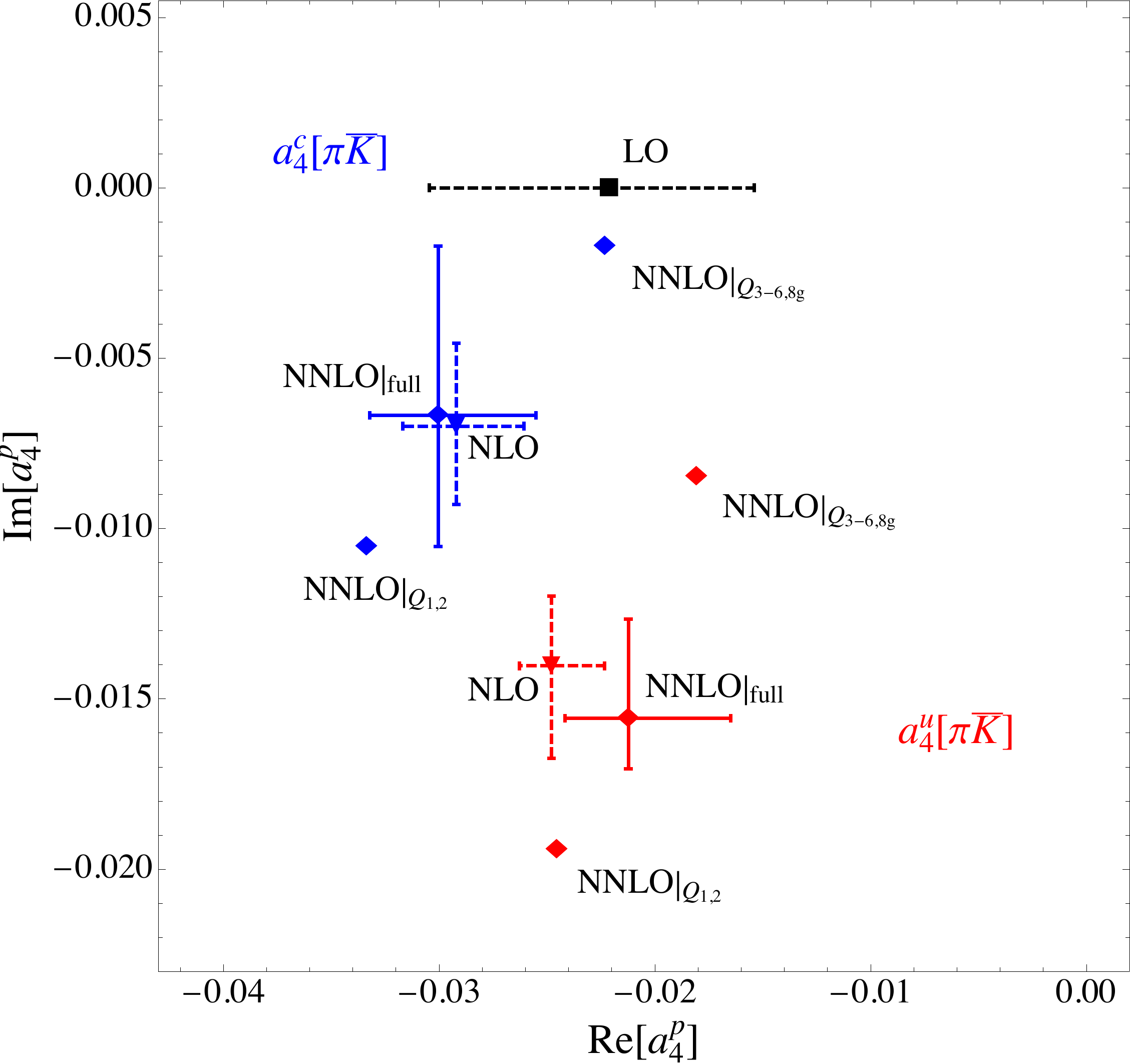}
\end{center}

\vspace*{-12pt}

\caption{Anatomy of QCD corrections to $a_4^{u,c}(\pi\bar K)$~\cite{Bell:2020qus}. The leading-order (LO) point is equal for both.\label{fig:penguinamplitudesNNLO}}
\end{figure}

The QCD factorization formula~\cite{Beneke:1999br,Beneke:2001ev,Beneke:2003zv} for charmless two-body nonleptonic $B$ decays into two pseudoscalar mesons $M_1$ and $M_2$,
\begin{eqnarray}
\label{factformula}
\langle M_1 M_2 | Q_i | \bar{B} \rangle & = &
i \,\frac{m_B^2}{4} \,\bigg\{F^{BM_1}(0)
\int_0^1 \! du \; T_{i}^I(u) \, f_{M_2}\phi_{M_2}(u)
+ (M_1\leftrightarrow M_2) \nnb \\
&& \hspace*{-2.1cm}
+ \,\int_0^\infty \! d\omega \int _0^1 \! du dv \;
T_{i}^{II}(\omega,v,u) \, f_B \phi_B(\omega)  \; 
f_{M_1}\phi_{M_1}(v) \;
f_{M_2} \phi_{M_2}(u) \bigg\} + \mathcal{O}\left(\frac{\Lambda_{\rm QCD}}{m_b}\right) \, ,
\end{eqnarray}
expresses the matrix element of an operator $Q_i$ from the effective weak Hamiltonian in terms of non-perturbative $B\to M$ form factors $F^{BM}(0)$, decay constants $f_M$, light-cone distribution amplitudes (LCDA) $\phi_{M}(u)$, and perturbatively calculable hard-scattering kernels $T_{i}^I(u)$, $T_{i}^{II}(\omega,v,u)$.
Since the latter have the structure $T_{i}^I = 1 + \mathcal{O}(\alpha_s)$ and $T_{i}^{II} = \mathcal{O}(\alpha_s)$, it is clear that the (strong) rescattering phases and hence the direct CP asymmetries are of ${\cal O}(\alpha_s)$, or of next-to-leading power $\mathcal{O}(\Lambda_{\rm QCD}/m_b)$. While the calculation of power corrections in QCDF remains challenging, the short-distance part at leading power has recently been completed at next-to-next-to-leading order (NNLO)~\cite{Bell:2020qus}, corresponding to ${\cal O}(\alpha_s^2)$. The calculation in the latter paper focused on the two-loop correction to the penguin amplitudes $a_4^u(M_1M_2)$ and $a_4^c(M_1M_2)$, which required the computation of more than a hundred two-loop diagrams involving two different scales ($u$ and $m_c^2/m_b^2$), and could only be achieved by applying sophisticated multi-loop techniques, in particular for the master integrals~\cite{Bell:2014zya}. In figure~\ref{fig:penguinamplitudesNNLO} we provide for $a_4^u(\pi\bar K)$ and $a_4^c(\pi\bar K)$ the anatomy of the QCD corrections at various orders~\cite{Bell:2020qus}. Recently, also QED corrections became available~\cite{Beneke:2020vnb,Beneke:2021jhp}.

The results for the amplitudes can be used to obtain the direct CP asymmetries at ${\cal O}(\alpha_s^2)$. In table~\ref{tab:CPAs} we collect the numerical values of direct CP asymmetries for a sample of penguin-dominated charmless $B \to PP$ channels at various perturbative orders\footnote{Note that the numbers in table~\ref{tab:CPAs} are from~\cite{Bell:2015koa}, where at NNLO only the contribution from current-current operators to $a_4^{u,c}$ was available. A comprehensive update using also more recent results is still pending.}.
The columns labelled ``NLO'' and ``NNLO'' give the respective results with all power-suppressed terms but the short-distance dominated scalar penguin amplitude set to zero. The column ``NNLO+LD'' adds the previously neglected long-distance (LD) terms back, whose main effect is from weak annihilation. The numbers in table~\ref{tab:CPAs} reveal that the perturbatvie NNLO corrections are in general not sizeable, as can be anticipated from figure~\ref{fig:penguinamplitudesNNLO}. The addition of the LD effects, which is done according to the scenario $S_4^\prime$ in~\cite{Beneke:2003zv}, has a large impact on both, the central values and the uncertainties, and in many cases spoils the precision achieved at leading power. There are, however, suitably chosen combinations which are robust against power corrections, for instance the CP asymmetry difference
\begin{equation}\delta(\pi \bar{K}) =
A_{\rm CP}(\pi^0 K^-) - A_{\rm CP}(\pi^+ K^-) \, ,
\end{equation}
which is still at the heart of the $B\to K\pi$ puzzle, or the asymmetry sum rule
\begin{equation}
\Delta(\pi \bar{K}) = A_{\rm CP}(\pi^+ K^-) +
\frac{\Gamma_{\pi^-\bar{K}^0}}{\Gamma_{\pi^+ K^-}} \,
A_{\rm CP}(\pi^- \bar{K}^0)
- \frac{2\Gamma_{\pi^0 K^-}}{\Gamma_{\pi^+ K^-}} \,
A_{\rm CP}(\pi^0 K^-)
- \frac{2 \Gamma_{\pi^0 \bar{K}^0}}{\Gamma_{\pi^+ K^-}} \,
A_{\rm CP}(\pi^0 \bar{K}^0)\, ,
\label{eq:sumrule}
\end{equation}
which is expected to be small \cite{Gronau:2005kz}. The numbers in table~\ref{tab:CPAs} confirm that power corrections are numerically much better under control for these quantities than for most of the direct CP asymmetries themselves. An updated value for $\Delta(\pi \bar{K})$ including QED corrections can be found in~\cite{Beneke:2020vnb}. On the experimental side one of the urgently missing pieces is the CP asymmetry in the $\pi^0\bar K^0$ channel, which constitutes a very important measurement at Belle~II. Recent progress in this direction was reported in~\cite{Belle-II:2021jvj}.

\begin{table*}[t]
\tabcolsep0.2cm
 \let\oldarraystretch=\arraystretch
 \renewcommand*{\arraystretch}{1.1}
\begin{center}\scalebox{0.85}{
\begin{tabular}{lcccc}
\hline \hline
&&&&\\[-0.3cm]
$f$  & ${\rm NLO}$ & ${\rm NNLO}$ &  ${\rm NNLO}+{\rm LD}$ & Exp \\
\hline
&&&&\\[-0.1cm]
$\pi^-\bar{K}^0$
& $\phantom{-}0.71_{\,-0.14\,-0.19}^{\,+0.13\,+0.21}$
& $\phantom{-}0.77_{\,-0.15\,-0.22}^{\,+0.14\,+0.23}$
& $\phantom{-}0.10_{\,-0.02\,-0.27}^{\,+0.02\,+1.24}$
& $-1.7\pm1.6$ \\[1em]
$\pi^0K^-$
& $\phantom{-}9.42_{\,-1.76\,-1.88}^{\,+1.77\,+1.87}$
& $10.18_{\,-1.90\,-2.62}^{\,+1.91\,+2.03}$
& $-1.17_{\,-0.22\,-\phantom{0}6.62}^{\,+0.22\,+20.00}$
& $\phantom{-}4.0\pm2.1$ \\[1em]
$\pi^+K^-$
& $\phantom{-}7.25_{\,-1.36\,-2.58}^{\,+1.36\,+2.13}$
& $\phantom{-}8.08_{\,-1.51\,-2.65}^{\,+1.52\,+2.52}$
& $-3.23_{\,-0.61\,-\phantom{0}3.36}^{\,+0.61\,+19.17}$
& $-8.2\pm0.6$ \\[1em]
$\pi^0\bar{K}^0$
& $-4.27_{\,-0.77\,-2.23}^{\,+0.83\,+1.48}$
& $-4.33_{\,-0.78\,-2.32}^{\,+0.84\,+3.29}$
& $-1.41_{\,-0.25\,-6.10}^{\,+0.27\,+5.54}$
& $\phantom{-1}1\pm10$ \\[1em]
$\delta(\pi \bar{K})$
& $\phantom{-}2.17_{\,-0.40\,-0.74}^{\,+0.40\,+1.39}$
& $\phantom{-}2.10_{\,-0.39\,-2.86}^{\,+0.39\,+1.40}$
& $\phantom{-}2.07_{\,-0.39\,-4.55}^{\,+0.39\,+2.76}$
& $12.2\pm 2.2$ \\[1em]
$\Delta(\pi \bar{K})$
& $-1.15_{\,-0.22\,-0.84}^{\,+0.21\,+0.55}$
& $-0.88_{\,-0.17\,-0.91}^{\,+0.16\,+1.31}$
& $-0.48_{\,-0.09\,-1.15}^{\,+0.09\,+1.09}$
& $-14\pm11$ \\[1em]
\hline
\end{tabular}}
\end{center}

\vspace*{-12pt}

\caption{Direct CP asymmetries in percent~\cite{Bell:2015koa}. Theoretical uncertainties are CKM and hadronic, respectively.
 \label{tab:CPAs}}
\end{table*}

\section{Puzzles in tree-level color-allowed decays}

For two-body colour-allowed nonleptonic tree-level decays such as $\bar B_{(s)}^0 \to D_{(s)}^{(\ast)+} L^-$ ($L$ being a light pseudoscalar meson) QCDF is expected to work very well: Both the colour-suppressed tree and the penguin amplitudes are absent, and effects from spectator scattering and weak annihilation are power suppressed~\cite{Beneke:2000ry}. Moreover, weak annihilation is absent if the decay is flavour-specific, i.e.\ if all final-state flavours are distinct as in $\bar{B}_s^0 \to D_s^{(\ast)+} \pi^-$ and $\bar{B}^0 \to D^{(\ast)+} K^-$ (but not in $\bar{B}^0 \to D^{(\ast)+} \pi^-$). Besides the branching fractions themselves we will consider the following ratios,
\begin{align}
\mathcal{R}^{P(V)}_{s/d}  &=  \frac{\mathcal{B}(\bar{B}_s^0 \to D_s^{(*)+} \pi^-)}{\mathcal{B}(\bar{B}^0 \to D^{(*)+} K^-)} \, , &
\mathcal R_s^{V/P} &= \frac{\mathcal B(\bar B_s^0\to D_s^{*+}\pi^-)}{\mathcal B(\bar B_s^0\to D_s^{+}\pi^-)}\, , &
\mathcal R_d^{V/P} &= \frac{\mathcal B(\bar {B}^0\to D^{*+}K^-)}{\mathcal B(\bar B^0\to D^{+}K^-)}\, . \label{eq:nonlepratios}
\end{align}
In the factorisation formula~\cite{Beneke:2000ry}
\begin{align}
    \bra{D_{(s)}^{(*)+} L^-} \mathcal{Q}_i \ket{\bar{B}_{(s)}^0}
        = & \sum_{j} F_j^{\bar{B}_{(s)} \to D_{(s)}^{(*)}}\!(M_L^2) \, \int_0^1 du \, T_{ij}(u) \phi_L(u)+ \order{\frac{\Lambda_\text{QCD}}{m_b}}
\end{align}
the hard functions $T_{ij}(u)$ are known to two loops~\cite{Huber:2016xod}, and the form factors have been examined in a recent precision study~\cite{Bordone:2019guc}. As expected, the bottleneck to a precision prediction are the power corrections. In~\cite{Bordone:2020gao}, power corrections from several effects were identified and their size estimated: higher twist effects to the light-meson LCDA, hard-collinear gluon emission from the spectator quark $q$ and from the heavy quarks $b$ and $c$, and soft-gluon exchange between the $B \to D$ and the light-meson system. The total size of the next-to-leading power compared to the leading-power contributions were conservatively estimated to be
below the percent level~\cite{Bordone:2020gao}, which supports the picture of these decays being very clean.

\begin{table*}[t]
    \centering
    \scalebox{0.90}{
    \begin{tabular}{l|c|c|c}
        \hline
                                                  & PDG                   & QCDF prediction              & discrepancy\\
        \hline
        $\mathcal{B}(\bar B_s^0\to D_s^+\pi^-)$   & $3.00\pm0.23$         & $4.42 \pm 0.21$              & $\sim 4 \sigma$ \\
        $\mathcal{B}(\bar B^0\to D^+K^-)$         & $0.186\pm0.020$       & $0.326 \pm 0.015$            & $\sim 5 \sigma$ \\
        $\mathcal{B}(\bar B_s^0\to D_s^{*+}\pi^-)$& $2.0\pm0.5$           & $4.3^{+0.9}_{-0.8}$          & $\sim 2 \sigma$ \\
        $\mathcal{B}(\bar B^0\to D^{*+}K^-)$      & $0.212\pm 0.015$      & $0.327^{+0.039}_{-0.034}$    & $\sim 3 \sigma$ \\
        \hline
        $\mathcal R_{s/d}^P$                      & $16.1\pm2.1$          & $13.5^{+0.6}_{-0.5}$         & $< 1 \sigma$    \\
        $\mathcal R_{s/d}^V$                      & $9.4\pm2.5$           & $13.1^{+2.3}_{-2.0}$         & $< 1 \sigma$    \\
        $\mathcal R_s^{V/P}$                      & $0.66\pm0.16$         & $0.97^{+0.20}_{-0.17}$       & $< 1 \sigma$    \\
        $\mathcal R_d^{V/P}$                      & $1.14\pm 0.15$        & $1.01 \pm 0.11$              & $< 1 \sigma$    \\
        \bottomrule
    \end{tabular}}
    \caption{Theory vs.\ experiment for flavour-specific, colour-allowed tree-level decays. Branching ratios are given in units of $10^{-3}$, their ratios are defined in eq.~\eqref{eq:nonlepratios}.\label{tab:resultsBR}}
\end{table*}

However, when comparing to the experimental measurements a puzzling pattern arises, which we summarize in table~\ref{tab:resultsBR}. The experimental values for the branching ratios are consistenly below the theoretical predictions, between $\sim 2\sigma$ and $\sim 5\sigma$ depending on the channel. The ratios of branching fractions are, on the other hand, in agreement within uncertainties. The numbers could be brought into agreement with a universal, non-factorizable contributions of ${\cal O}(-15-20\%)$ on the amplitude level. At the moment, it is unclear where this contribution could arise. QED corrections were studied in~\cite{Beneke:2021jhp}, with the outcome that they ease the tension but are too small to explain the discrepancy. Rescattering effects were considered in~\cite{Endo:2021ifc} and also found to be too small. Also the input parameters can be considered reliable. Issues on the experimental side are also unlikely since the decays have a large branching fraction and only charged particles in the final state. Moreover, a recent measurement from Belle~\cite{Belle:2021udv} confirms earlier findings. Effects of physics from beyond the Standard Model (BSM) have also been investigated in a couple of recent papers.  In~\cite{Iguro:2020ndk} it was found that the tension can be partially explained by a left-handed $W^\prime$-model, while still being compatible with other flavour and collider bounds. New tensor structures were analysed in~\cite{Cai:2021mlt}, some of them can explain the data at the $1\sigma$-level. In the same reference also a model-dependent analysis, e.g.\ with a colourless charged scalar, was carried out.
In~\cite{Bordone:2021cca} a combination with dijet searches was performed, where mediators with various $SU(3) \times SU(2) \times U(1)$ quantum numbers were considered. For a good portion of the scenarios considered there, it was pointed out that the parameter space to explain the nonleptonic tree-level puzzle is to a large extent already ruled out by dijet searches. Other recent studies that combine tree-level nonleptonic decays with lifetimes~\cite{Lenz:2019lvd}, with $B_s^0 \to D_s^{\mp} K^\pm$ decays~\cite{Fleischer:2021cct,Fleischer:2021cwb}, and with CP violation in the mixing and decay of $B^0_{(s)}$ mesons~\cite{Gershon:2021pnc} are also available.

\begin{figure}
\begin{center}
\includegraphics[width=0.49\textwidth]{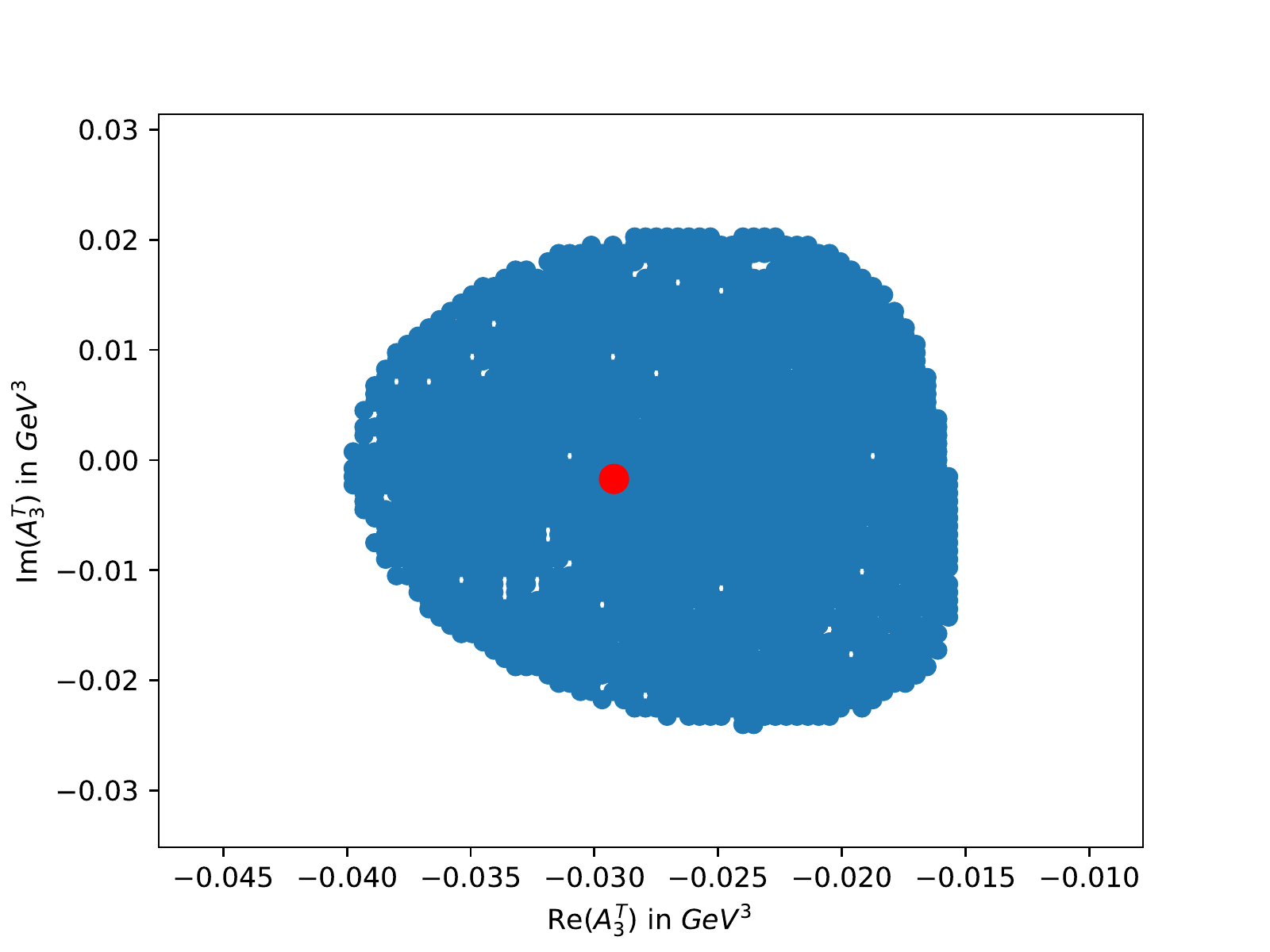}
\includegraphics[width=0.49\textwidth]{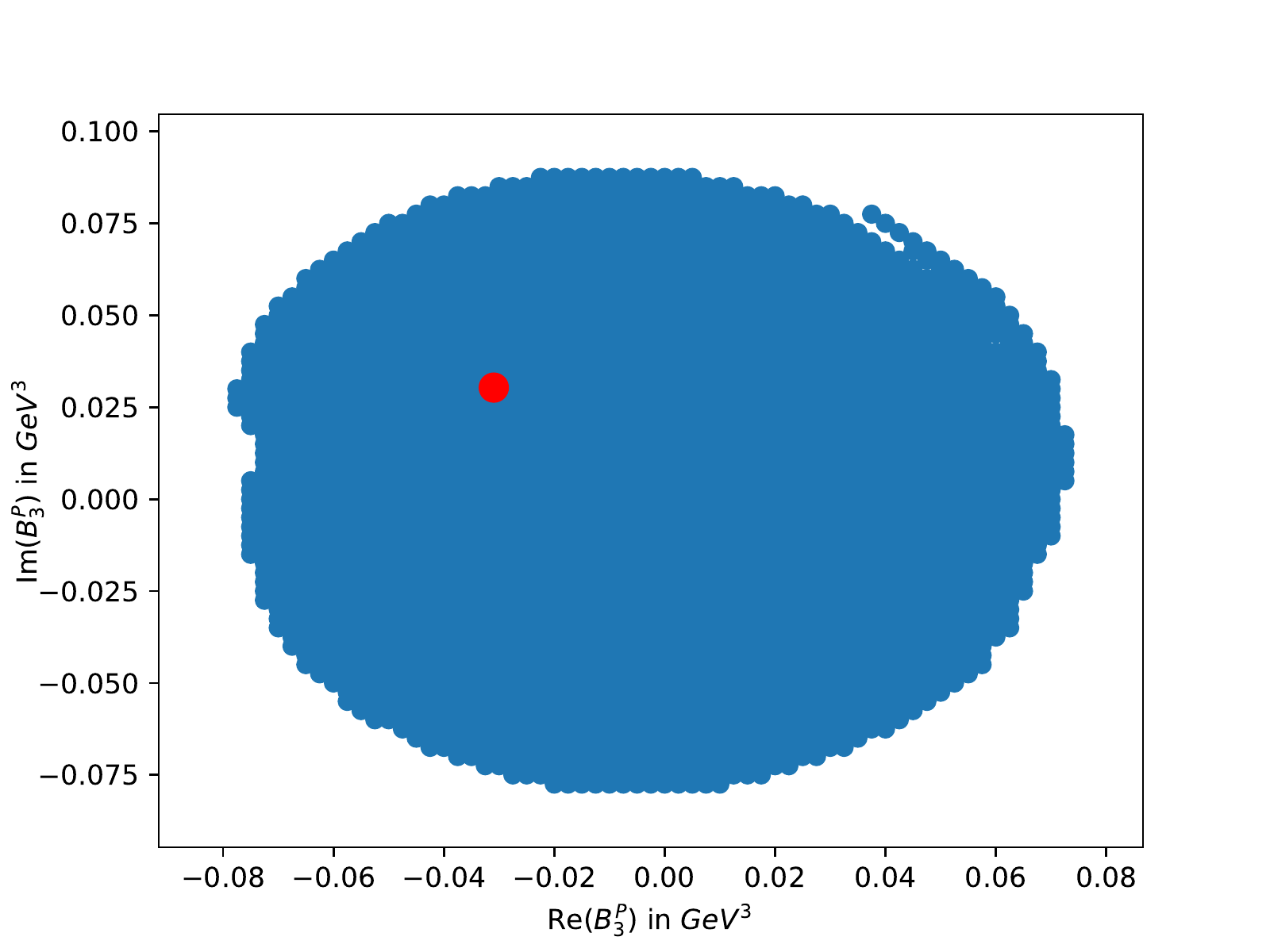}
\end{center}

\vspace*{-12pt}

\caption{Two sample amplitudes determined from the $SU(3)$-irreducible fit. The red point denotes the best-fit point, the blue area the $1\sigma$ confidence region.\label{fig:SU3amplitudes}}
\end{figure}

\section{Combining QCD factorisation and flavour symmetries}

In a recent work~\cite{Huber:2021cgk} we estimate the potential size of the weak-annihilation amplitudes in QCDF in a data-driven approach combined with the $SU(3)$ flavour symmetry of the light quarks. To this end we use the linear relations between the so-called topological and $SU(3)$-invariant description of the decay amplitude~\cite{He:2018php}, and determine the $SU(3)$-invariant amplitudes through a $\chi^2$-fit, for which we use experimental input for branching fractions (23 measurements and 6 upper bounds) and CP asymmetries (17 measurements and one upper bound) from $B \to PP$ decays. The fit parameters are made up of 20 complex amplitudes (10 for tree and 10 for penguins), of which two complex amplitudes and one overall phase can be absorbed. Together with the $\eta~-~\eta^\prime$ mixing angle, we therefore fit for 36 real parameters. The best-fit point is determined from $10^9$ randomly generated points in our 36-dimensional space, with some refinements for which we refer to~\cite{Huber:2021cgk}. The uncertainties are determined through a likelihood ratio test, and the $p$ value is determined from Wilk's theorem with two degrees of freedom. In figure~\ref{fig:SU3amplitudes} we show the fit result for two sample $SU(3)$-irreducible amplitudes. To assess the physical consequences of the fit, the results for the amplitudes are translated back to branching fractions and direct CP asymmetries for more than 30 $B \to PP$ channels. Quantitatively, the goodness of the fit is reflected by $\chi^2/d.o.f = 0.851$. Hence, the vast majority of the numbers between theory and experiment are in agreement, though with still sizeable uncertainties in certain cases. Moreover, some of our numbers represent predictions for yet unmeasured channels, in particular those from $B^0_s$ decays and those with $\eta^{(\prime)}$ in the final state.

\begin{figure}
\begin{center}
\includegraphics[width=0.49\textwidth]{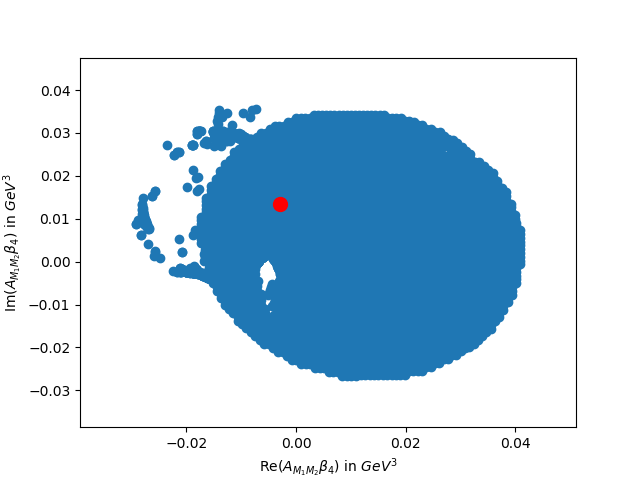}
\includegraphics[width=0.49\textwidth]{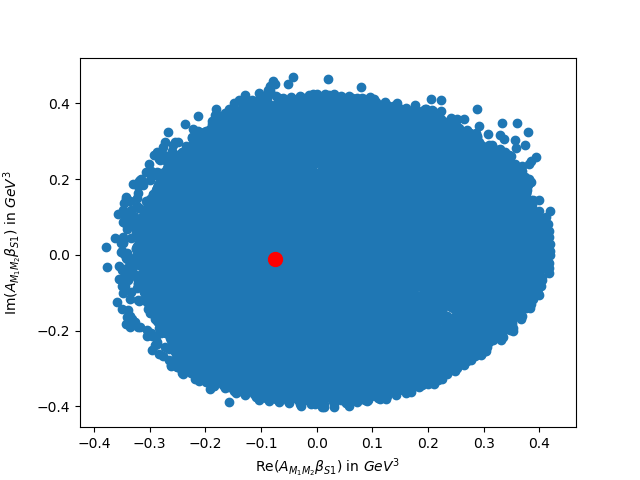}
\end{center}

\vspace*{-12pt}

\caption{Fit results for two sample annihilation amplitudes. The colour coding is the same as in figure~\ref{fig:SU3amplitudes}.\label{fig:annihilationamplitudes}}
\end{figure}

We then investigate the connection between the $SU(3)$-irreducible representation and QCDF. We establish the transformation rules between the topological description and QCDF, and show that the number of independent complex amplitudes equals 18 in both approaches (see also~\cite{beneketalk}), and that the relation is again linear. Together with the relations in~\cite{He:2018php} this establishes the transformation between the $SU(3)$-irreducible and the QCDF amplitudes, and hence allows for the
translation of the $SU(3)$-fit results into constraints on QCDF amplitudes. In particular, the fit gives a quantitative estimate of the size of the annihilation amplitudes as dictated by data (see also~\cite{Bobeth:2014rra}). Our main finding is that the most constrained weak annihilation amplitudes are below $0.04$, see left panel of figure~\ref{fig:annihilationamplitudes}. However, values up to $\sim 0.3$ are allowed in certain cases (see right panel of figure~\ref{fig:annihilationamplitudes}), which can to a large part be attributed to sizeable uncertainties in several of the experimental input parameters. Also effects of $SU(3)$ breaking are briefly addressed in~\cite{Huber:2021cgk}. However, this important topic certainly deserves further dedicated studies in the future.

\section{Conclusion}

Despite all the progress in nonleptonic $B_{(s)}$ decays, the field is eagerly awaiting further improvements. On the experimental side a key input to the asymmetry sum rule is the measurement of the direct CP asymmetry in $\bar B^0 \to \pi^0\bar K^0$ (for a first step in this direction see~\cite{Belle-II:2021jvj}). Moreover, measurements of branching ratios and direct CP asymmetries in $B_{(s)}$ decays to $\eta^{(\prime)}$ will yield a more comprehensive picture in the comparison between theory and experiment. On the theoretical side it is indispensable to get power corrections and flavour symmetry breaking under better control. For a very recent work see~\cite{Lu:2022fgz}.

\section*{Acknowledgment}

I would like to thank the organisers of CKM 2021 for creating a lively and inspiring atmosphere despite the difficult pandemic circumstances. Moreover, I would like to thank Gilberto Tetlalmatzi-Xolocotzi for feedback on the manuscript, and all co-authors of the projects reported here for most enjoyable collaborations. 
This work was supported by the Deutsche Forschungsgemeinschaft (DFG, German Research Foundation) under grant 396021762~--~TRR 257 ``Particle Physics Phenomenology after the Higgs Discovery''.


\begin{thebibliography}{10}

\bibitem{Keum:2000ph}
Y.-Y.~Keum, H.-n.~Li and A.I.~Sanda, \emph{{Fat penguins and imaginary penguins
  in perturbative QCD}},
  \href{https://doi.org/10.1016/S0370-2693(01)00247-7}{\emph{Phys. Lett. B}
  {\bfseries 504} (2001) 6}
  [\href{https://arxiv.org/abs/hep-ph/0004004}{{\ttfamily hep-ph/0004004}}].

\bibitem{Lu:2000em}
C.-D.~Lu, K.~Ukai and M.-Z.~Yang, \emph{{Branching ratio and CP violation of $B
  \to \pi \pi$ decays in perturbative QCD approach}},
  \href{https://doi.org/10.1103/PhysRevD.63.074009}{\emph{Phys. Rev. D}
  {\bfseries 63} (2001) 074009}
  [\href{https://arxiv.org/abs/hep-ph/0004213}{{\ttfamily hep-ph/0004213}}].

\bibitem{Beneke:1999br}
M.~Beneke, G.~Buchalla, M.~Neubert and C.T.~Sachrajda, \emph{{QCD factorization
  for $B \to \pi \pi$ decays: Strong phases and CP violation in the heavy quark
  limit}}, \href{https://doi.org/10.1103/PhysRevLett.83.1914}{\emph{Phys. Rev.
  Lett.} {\bfseries 83} (1999) 1914}
  [\href{https://arxiv.org/abs/hep-ph/9905312}{{\ttfamily hep-ph/9905312}}].

\bibitem{Beneke:2000ry}
M.~Beneke, G.~Buchalla, M.~Neubert and C.T.~Sachrajda, \emph{{QCD factorization
  for exclusive, nonleptonic B meson decays: General arguments and the case of
  heavy light final states}},
  \href{https://doi.org/10.1016/S0550-3213(00)00559-9}{\emph{Nucl. Phys. B}
  {\bfseries 591} (2000) 313}
  [\href{https://arxiv.org/abs/hep-ph/0006124}{{\ttfamily hep-ph/0006124}}].

\bibitem{Beneke:2001ev}
M.~Beneke, G.~Buchalla, M.~Neubert and C.T.~Sachrajda, \emph{{QCD factorization
  in $B \to \pi K$, $\pi \pi$ decays and extraction of Wolfenstein
  parameters}},
  \href{https://doi.org/10.1016/S0550-3213(01)00251-6}{\emph{Nucl. Phys. B}
  {\bfseries 606} (2001) 245}
  [\href{https://arxiv.org/abs/hep-ph/0104110}{{\ttfamily hep-ph/0104110}}].

\bibitem{Khodjamirian:2005wn}
A.~Khodjamirian, T.~Mannel, M.~Melcher and B.~Melic, \emph{{Annihilation
  effects in $B \to \pi \pi$ from QCD light-cone sum rules}},
  \href{https://doi.org/10.1103/PhysRevD.72.094012}{\emph{Phys. Rev. D}
  {\bfseries 72} (2005) 094012}
  [\href{https://arxiv.org/abs/hep-ph/0509049}{{\ttfamily hep-ph/0509049}}].

\bibitem{Zeppenfeld:1980ex}
D.~Zeppenfeld, \emph{{SU(3) Relations for B Meson Decays}},
  \href{https://doi.org/10.1007/BF01429835}{\emph{Z. Phys. C} {\bfseries 8}
  (1981) 77}.

\bibitem{Savage:1989ub}
M.J.~Savage and M.B.~Wise, \emph{{SU(3) Predictions for Nonleptonic B Meson
  Decays}}, \href{https://doi.org/10.1103/PhysRevD.39.3346}{\emph{Phys. Rev. D}
  {\bfseries 39} (1989) 3346}.

\bibitem{Gronau:1990ka}
M.~Gronau and D.~London, \emph{{Isospin analysis of CP asymmetries in B
  decays}}, \href{https://doi.org/10.1103/PhysRevLett.65.3381}{\emph{Phys. Rev.
  Lett.} {\bfseries 65} (1990) 3381}.

\bibitem{Krankl:2015fha}
S.~Kr\"ankl, T.~Mannel and J.~Virto, \emph{{Three-body non-leptonic B decays
  and QCD factorization}},
  \href{https://doi.org/10.1016/j.nuclphysb.2015.08.004}{\emph{Nucl. Phys. B}
  {\bfseries 899} (2015) 247}
  [\href{https://arxiv.org/abs/1505.04111}{{\ttfamily 1505.04111}}].

\bibitem{Klein:2017xti}
R.~Klein, T.~Mannel, J.~Virto and K.K.~Vos, \emph{{CP Violation in Multibody
  $B$ Decays from QCD Factorization}},
  \href{https://doi.org/10.1007/JHEP10(2017)117}{\emph{JHEP} {\bfseries 10}
  (2017) 117} [\href{https://arxiv.org/abs/1708.02047}{{\ttfamily
  1708.02047}}].

\bibitem{Huber:2020pqb}
T.~Huber, J.~Virto and K.K.~Vos, \emph{{Three-Body Non-Leptonic Heavy-to-heavy
  $B$ Decays at NNLO in QCD}},
  \href{https://doi.org/10.1007/JHEP11(2020)103}{\emph{JHEP} {\bfseries 11}
  (2020) 103} [\href{https://arxiv.org/abs/2007.08881}{{\ttfamily
  2007.08881}}].

\bibitem{Li:2012cfa}
H.-n.~Li, C.-D.~Lu and F.-S.~Yu, \emph{{Branching ratios and direct CP
  asymmetries in $D\to PP$ decays}},
  \href{https://doi.org/10.1103/PhysRevD.86.036012}{\emph{Phys. Rev. D}
  {\bfseries 86} (2012) 036012}
  [\href{https://arxiv.org/abs/1203.3120}{{\ttfamily 1203.3120}}].

\bibitem{Qin:2013tje}
Q.~Qin, H.-n.~Li, C.-D.~L\"u and F.-S.~Yu, \emph{{Branching ratios and direct
  CP asymmetries in $D\to PV$ decays}},
  \href{https://doi.org/10.1103/PhysRevD.89.054006}{\emph{Phys. Rev. D}
  {\bfseries 89} (2014) 054006}
  [\href{https://arxiv.org/abs/1305.7021}{{\ttfamily 1305.7021}}].

\bibitem{Zhou:2015jba}
S.-H.~Zhou, Y.-B.~Wei, Q.~Qin, Y.~Li, F.-S.~Yu and C.-D.~Lu, \emph{{Analysis of
  Two-body Charmed $B$ Meson Decays in Factorization-Assisted
  Topological-Amplitude Approach}},
  \href{https://doi.org/10.1103/PhysRevD.92.094016}{\emph{Phys. Rev. D}
  {\bfseries 92} (2015) 094016}
  [\href{https://arxiv.org/abs/1509.04060}{{\ttfamily 1509.04060}}].

\bibitem{Wang:2017hxe}
C.~Wang, Q.-A.~Zhang, Y.~Li and C.-D.~Lu, \emph{{Charmless $B_{(s)}\to VV$
  Decays in Factorization-Assisted Topological-Amplitude Approach}},
  \href{https://doi.org/10.1140/epjc/s10052-017-4889-3}{\emph{Eur. Phys. J. C}
  {\bfseries 77} (2017) 333}
  [\href{https://arxiv.org/abs/1701.01300}{{\ttfamily 1701.01300}}].

\bibitem{Jiang:2017zwr}
H.-Y.~Jiang, F.-S.~Yu, Q.~Qin, H.-n.~Li and C.-D.~L\"u,
  \emph{{$D^0$-$\overline{D}^0$ mixing parameter $y$ in the
  factorization-assisted topological-amplitude approach}},
  \href{https://doi.org/10.1088/1674-1137/42/6/063101}{\emph{Chin. Phys. C}
  {\bfseries 42} (2018) 063101}
  [\href{https://arxiv.org/abs/1705.07335}{{\ttfamily 1705.07335}}].

\bibitem{Gronau:1995hn}
M.~Gronau, O.F.~Hernandez, D.~London and J.L.~Rosner, \emph{{Electroweak
  penguins and two-body B decays}},
  \href{https://doi.org/10.1103/PhysRevD.52.6374}{\emph{Phys. Rev. D}
  {\bfseries 52} (1995) 6374}
  [\href{https://arxiv.org/abs/hep-ph/9504327}{{\ttfamily hep-ph/9504327}}].

\bibitem{Descotes-Genon:2006spp}
S.~Descotes-Genon, J.~Matias and J.~Virto, \emph{{Exploring $B_{d,s} \to KK$
  decays through flavour symmetries and QCD-factorisation}},
  \href{https://doi.org/10.1103/PhysRevLett.97.061801}{\emph{Phys. Rev. Lett.}
  {\bfseries 97} (2006) 061801}
  [\href{https://arxiv.org/abs/hep-ph/0603239}{{\ttfamily hep-ph/0603239}}].

\bibitem{Cheng:2011qh}
H.-Y.~Cheng and S.~Oh, \emph{{Flavor SU(3) symmetry and QCD factorization in $B
  \to PP$ and $PV$ decays}},
  \href{https://doi.org/10.1007/JHEP09(2011)024}{\emph{JHEP} {\bfseries 09}
  (2011) 024} [\href{https://arxiv.org/abs/1104.4144}{{\ttfamily 1104.4144}}].

\bibitem{Hsiao:2015iiu}
Y.-K.~Hsiao, C.-F.~Chang and X.-G.~He, \emph{{A global $SU(3)/U(3)$ flavor
  symmetry analysis for $B\to PP$ with $\eta-\eta'$ Mixing}},
  \href{https://doi.org/10.1103/PhysRevD.93.114002}{\emph{Phys. Rev. D}
  {\bfseries 93} (2016) 114002}
  [\href{https://arxiv.org/abs/1512.09223}{{\ttfamily 1512.09223}}].

\bibitem{Huber:2021cgk}
T.~Huber and G.~Tetlalmatzi-Xolocotzi, \emph{{Estimating QCD-factorization
  amplitudes through SU(3) symmetry in $B\rightarrow P P$ decays}},
  \href{https://doi.org/10.1140/epjc/s10052-022-10068-8}{\emph{Eur. Phys. J. C}
  {\bfseries 82} (2022) 210}
  [\href{https://arxiv.org/abs/2111.06418}{{\ttfamily 2111.06418}}].

\bibitem{Bhattacharya:2014eca}
B.~Bhattacharya, M.~Gronau, M.~Imbeault, D.~London and J.L.~Rosner,
  \emph{{Charmless B$\to$PPP decays: The fully-symmetric final state}},
  \href{https://doi.org/10.1103/PhysRevD.89.074043}{\emph{Phys. Rev. D}
  {\bfseries 89} (2014) 074043}
  [\href{https://arxiv.org/abs/1402.2909}{{\ttfamily 1402.2909}}].

\bibitem{Bhattacharya:2015uua}
B.~Bhattacharya and D.~London, \emph{{Using U spin to extract $\gamma$ from
  charmless $B \to PPP$ decays}},
  \href{https://doi.org/10.1007/JHEP04(2015)154}{\emph{JHEP} {\bfseries 04}
  (2015) 154} [\href{https://arxiv.org/abs/1503.00737}{{\ttfamily
  1503.00737}}].

\bibitem{Bediaga:2021okg}
I.~Bediaga, T.~Frederico, P.C.~Magalhaes and D.T.~Machado, \emph{{Global CP
  asymmetries in charmless three-body B decays with final state interactions}},
  \href{https://doi.org/10.1016/j.physletb.2021.136824}{\emph{Phys. Lett. B}
  {\bfseries 824} (2022) 136824}
  [\href{https://arxiv.org/abs/2109.01625}{{\ttfamily 2109.01625}}].

\bibitem{Beneke:2003zv}
M.~Beneke and M.~Neubert, \emph{{QCD factorization for $B \to PP$ and $B \to
  PV$ decays}},
  \href{https://doi.org/10.1016/j.nuclphysb.2003.09.026}{\emph{Nucl. Phys. B}
  {\bfseries 675} (2003) 333}
  [\href{https://arxiv.org/abs/hep-ph/0308039}{{\ttfamily hep-ph/0308039}}].

\bibitem{Bell:2020qus}
G.~Bell, M.~Beneke, T.~Huber and X.-Q.~Li, \emph{{Two-loop non-leptonic penguin
  amplitude in QCD factorization}},
  \href{https://doi.org/10.1007/JHEP04(2020)055}{\emph{JHEP} {\bfseries 04}
  (2020) 055} [\href{https://arxiv.org/abs/2002.03262}{{\ttfamily
  2002.03262}}].

\bibitem{Bell:2014zya}
G.~Bell and T.~Huber, \emph{{Master integrals for the two-loop penguin
  contribution in non-leptonic B-decays}},
  \href{https://doi.org/10.1007/JHEP12(2014)129}{\emph{JHEP} {\bfseries 12}
  (2014) 129} [\href{https://arxiv.org/abs/1410.2804}{{\ttfamily 1410.2804}}].

\bibitem{Beneke:2020vnb}
M.~Beneke, P.~B\"oer, J.-N.~Toelstede and K.K.~Vos, \emph{{QED factorization of
  non-leptonic $B$ decays}},
  \href{https://doi.org/10.1007/JHEP11(2020)081}{\emph{JHEP} {\bfseries 11}
  (2020) 081} [\href{https://arxiv.org/abs/2008.10615}{{\ttfamily
  2008.10615}}].

\bibitem{Beneke:2021jhp}
M.~Beneke, P.~B\"oer, G.~Finauri and K.K.~Vos, \emph{{QED factorization of
  two-body non-leptonic and semi-leptonic B to charm decays}},
  \href{https://doi.org/10.1007/JHEP10(2021)223}{\emph{JHEP} {\bfseries 10}
  (2021) 223} [\href{https://arxiv.org/abs/2107.03819}{{\ttfamily
  2107.03819}}].

\bibitem{Bell:2015koa}
G.~Bell, M.~Beneke, T.~Huber and X.-Q.~Li, \emph{{Two-loop current-current
  operator contribution to the non-leptonic QCD penguin amplitude}},
  \href{https://doi.org/10.1016/j.physletb.2015.09.037}{\emph{Phys. Lett.}
  {\bfseries B750} (2015) 348}
  [\href{https://arxiv.org/abs/1507.03700}{{\ttfamily 1507.03700}}].

\bibitem{Gronau:2005kz}
M.~Gronau, \emph{{A Precise sum rule among four $B \to K \pi$ CP asymmetries}},
  \href{https://doi.org/10.1016/j.physletb.2005.09.014}{\emph{Phys. Lett. B}
  {\bfseries 627} (2005) 82}
  [\href{https://arxiv.org/abs/hep-ph/0508047}{{\ttfamily hep-ph/0508047}}].

\bibitem{Belle-II:2021jvj}
{\scshape Belle-II} collaboration, \emph{{First search for direct
  $CP$-violating asymmetry in $B^0 \to K^0 \pi^0$ decays at Belle II}},
  \href{https://arxiv.org/abs/2104.14871}{{\ttfamily 2104.14871}}.

\bibitem{Huber:2016xod}
T.~Huber, S.~Kr\"ankl and X.-Q.~Li, \emph{{Two-body non-leptonic heavy-to-heavy
  decays at NNLO in QCD factorization}},
  \href{https://doi.org/10.1007/JHEP09(2016)112}{\emph{JHEP} {\bfseries 09}
  (2016) 112} [\href{https://arxiv.org/abs/1606.02888}{{\ttfamily
  1606.02888}}].

\bibitem{Bordone:2019guc}
M.~Bordone, N.~Gubernari, D.~van Dyk and M.~Jung, \emph{{Heavy-Quark expansion
  for ${{\bar{B}}_s\rightarrow D^{(*)}_s}$ form factors and unitarity bounds
  beyond the ${SU(3)_F}$ limit}},
  \href{https://doi.org/10.1140/epjc/s10052-020-7850-9}{\emph{Eur. Phys. J. C}
  {\bfseries 80} (2020) 347}
  [\href{https://arxiv.org/abs/1912.09335}{{\ttfamily 1912.09335}}].

\bibitem{Bordone:2020gao}
M.~Bordone, N.~Gubernari, T.~Huber, M.~Jung and D.~van Dyk, \emph{{A puzzle in
  $\bar{B}_{(s)}^0 \to D_{(s)}^{(\ast)+} \lbrace \pi^-, K^-\rbrace$ decays and
  extraction of the $f_s/f_d$ fragmentation fraction}},
  \href{https://doi.org/10.1140/epjc/s10052-020-08512-8}{\emph{Eur. Phys. J. C}
  {\bfseries 80} (2020) 951}
  [\href{https://arxiv.org/abs/2007.10338}{{\ttfamily 2007.10338}}].

\bibitem{Endo:2021ifc}
M.~Endo, S.~Iguro and S.~Mishima, \emph{{Revisiting rescattering contributions
  to $ \overline{B} _{(s)}$ \textrightarrow{} $ {D}_{(s)}^{\left(\ast \right)}M
  $ decays}}, \href{https://doi.org/10.1007/JHEP01(2022)147}{\emph{JHEP}
  {\bfseries 01} (2022) 147}
  [\href{https://arxiv.org/abs/2109.10811}{{\ttfamily 2109.10811}}].

\bibitem{Belle:2021udv}
{\scshape Belle} collaboration, \emph{{Study of $\overline{B}{}^0\rightarrow
  D^{+}h^{-} (h=K/\pi)$ decays at Belle}},
  \href{https://doi.org/10.1103/PhysRevD.105.012003}{\emph{Phys. Rev. D}
  {\bfseries 105} (2022) 012003}
  [\href{https://arxiv.org/abs/2111.04978}{{\ttfamily 2111.04978}}].

\bibitem{Iguro:2020ndk}
S.~Iguro and T.~Kitahara, \emph{{Implications for new physics from a novel
  puzzle in $\bar{B}_{(s)}^0 \to D^{(\ast)+}_{(s)} \lbrace \pi^-, K^- \rbrace$
  decays}}, \href{https://doi.org/10.1103/PhysRevD.102.071701}{\emph{Phys. Rev.
  D} {\bfseries 102} (2020) 071701}
  [\href{https://arxiv.org/abs/2008.01086}{{\ttfamily 2008.01086}}].

\bibitem{Cai:2021mlt}
F.-M.~Cai, W.-J.~Deng, X.-Q.~Li and Y.-D.~Yang, \emph{{Probing new physics in
  class-I B-meson decays into heavy-light final states}},
  \href{https://doi.org/10.1007/JHEP10(2021)235}{\emph{JHEP} {\bfseries 10}
  (2021) 235} [\href{https://arxiv.org/abs/2103.04138}{{\ttfamily
  2103.04138}}].

\bibitem{Bordone:2021cca}
M.~Bordone, A.~Greljo and D.~Marzocca, \emph{{Exploiting dijet resonance
  searches for flavor physics}},
  \href{https://doi.org/10.1007/JHEP08(2021)036}{\emph{JHEP} {\bfseries 08}
  (2021) 036} [\href{https://arxiv.org/abs/2103.10332}{{\ttfamily
  2103.10332}}].

\bibitem{Lenz:2019lvd}
A.~Lenz and G.~Tetlalmatzi-Xolocotzi, \emph{{Model-independent bounds on new
  physics effects in non-leptonic tree-level decays of B-mesons}},
  \href{https://doi.org/10.1007/JHEP07(2020)177}{\emph{JHEP} {\bfseries 07}
  (2020) 177} [\href{https://arxiv.org/abs/1912.07621}{{\ttfamily
  1912.07621}}].

\bibitem{Fleischer:2021cct}
R.~Fleischer and E.~Malami, \emph{{Using $B^0_s\to D_s^\mp K^\pm$ Decays as a
  Portal to New Physics}},  \href{https://arxiv.org/abs/2109.04950}{{\ttfamily
  2109.04950}}.

\bibitem{Fleischer:2021cwb}
R.~Fleischer and E.~Malami, \emph{{Revealing New Physics in $B^0_s\to D_s^\mp
  K^\pm$ Decays}},  \href{https://arxiv.org/abs/2110.04240}{{\ttfamily
  2110.04240}}.

\bibitem{Gershon:2021pnc}
T.~Gershon, A.~Lenz, A.V.~Rusov and N.~Skidmore, \emph{{Testing the Standard
  Model with CP-asymmetries in flavour-specific non-leptonic decays}},
  \href{https://arxiv.org/abs/2111.04478}{{\ttfamily 2111.04478}}.

\bibitem{He:2018php}
X.-G.~He and W.~Wang, \emph{{Flavor SU(3) Topological Diagram and Irreducible
  Representation Amplitudes for Heavy Meson Charmless Hadronic Decays: Mismatch
  and Equivalence}},
  \href{https://doi.org/10.1088/1674-1137/42/10/103108}{\emph{Chin. Phys. C}
  {\bfseries 42} (2018) 103108}
  [\href{https://arxiv.org/abs/1803.04227}{{\ttfamily 1803.04227}}].

\bibitem{beneketalk}
M.~Beneke, ``Theory of two-body non-leptonic $b$ decays.'' 2019,
  https://indico.mitp.uni-mainz.de/event/177/.

\bibitem{Bobeth:2014rra}
C.~Bobeth, M.~Gorbahn and S.~Vickers, \emph{{Weak annihilation and new physics
  in charmless $B \to M M$ decays}},
  \href{https://doi.org/10.1140/epjc/s10052-015-3535-1}{\emph{Eur. Phys. J. C}
  {\bfseries 75} (2015) 340} [\href{https://arxiv.org/abs/1409.3252}{{\ttfamily
  1409.3252}}].

\bibitem{Lu:2022fgz}
C.-D.~L\"u, Y.-L.~Shen, C.~Wang and Y.-M.~Wang, \emph{{Enhanced
  Next-to-Leading-Order Corrections to Weak Annihilation $B$-Meson Decays}},
  \href{https://arxiv.org/abs/2202.08073}{{\ttfamily 2202.08073}}.

\end{thebibliography}

\providecommand{\href}[2]{#2}\begingroup\raggedright\endgroup

\end{document}